\documentclass[sigconf]{acmart}
\AtBeginDocument{%
  }

\usepackage{tabularx}
\usepackage{rotating}
\usepackage{xcolor}
\usepackage{multicol}
\usepackage{url}

\setcopyright{none}
\acmDOI{}

\acmConference[LIMITS '26]{12th Workshop on Computing within Limitsl}{June 23--25,
  2026}{Online}

\acmISBN{}
\setcopyright{cc}
\setcctype[4.0]{by}

\begin{document}

\title[Pushing the Limits]{Pushing the Limits: A Framework to Reform Institutional Ethics Review of Environmentally-Impactful Computing Research}

\author{Nicolas Gold}
\orcid{0000-0002-2195-5995}
\affiliation{%
  \institution{UCL Computer Science\\ University College London}
  \city{London}
  \country{UK}
}
\email{n.gold@ucl.ac.uk}

\author{Ross Purves}
\affiliation{%
  \institution{Institute of Education\\ University College London}
 \city{London}
 \country{UK}}
\email{r.purves@ucl.ac.uk}
\orcid{0000-0003-4301-7024}

\renewcommand{\shortauthors}{Gold and Purves}

\begin{abstract}
Computationally-intensive research (CIR) takes place on a wide variety of topics including AI.  Its environmental impact is potentially significant yet it does not always fall clearly within the scope of organisational ethics review policy on its own merits.  Many academic institutions have ethics oversight bodies (e.g. Research Ethics Committees or Institutional Review Boards) that occupy a potentially powerful position to encourage recognition of these issues and seek reflexive practice in researchers.  However, policies are often poorly-defined in respect of environmental issues and thus research is not reviewed, reviewers have little guidance for legitimate critique, and researchers are not challenged to consider planetary limits on computing resources and the interaction of these with their research.  This paper aims to address these problems by proposing scoping criteria for institutional ethics policy to bring CIR within the scope of ethics review on its own merits, framing evidential criteria for reviewers to apply in ethics review, and presenting a method by which CIR researchers can reflect on their proposed research in relation to environmental factors, and assess its potential value in the light of planetary limits.  
\end{abstract}


\maketitle

\section{Introduction}
\label{lab:introduction}
Computationally-intensive research (hereafter, CIR) takes place in academic and commercial settings at ever-increasing scale.  The term `computationally-intensive research' is used here to capture a wide variety of data science, machine learning, and other techniques that are compute-resource intensive (e.g. bioinformatics~\cite{grealeyCarbonFootprintBioinformatics2022}).  Whilst the popular term `AI' is often applied to these (and indeed, there are many ethical issues associated with such technology, e.g. see Floridi et al.~\cite{floridiAI4PeopleAnEthicalFramework2018}), environmental issues are not exclusive to `AI' technology but extend to any large-scale computational activity of which `AI' applications (e.g., Large Language Models (LLMs)) are but one example.  Operationalising the ethics review of CIR to consider environmental issues and planetary limits is challenging in scope, governance, and practice.  We address this challenge here.

CIR may not involve solely the creation of new algorithms and analytical approaches that directly involve heavy computation; a large and rapidly increasing volume of non computer-science research involves the use of LLMs as part of analytical tool-chains and methods, in the production of research software, and as objects of study in their own right.  Given the environmental costs of training and operating LLMs, non computer-science research that involves them might also be considered computationally-intensive.  

CIR frequently rests on the problematic underlying assumption identified by Nardi et al.~\cite{ComputingLimitsCommunications2018} and others in the LIMITS community, that exponential growth can continue indefinitely.
It increasingly relies on apparently de-materialised cloud resources and/or opaque commercial services that do not provide sufficient transparency for researchers to consider the non-local environmental costs of their methodological choices.  Accordingly, in the remainder of this paper we take a LIMITS-oriented perspective (after Nardi et al.~\cite{ComputingLimitsCommunications2018}) that there are finite limits to planetary resources, that growth cannot continue indefinitely (indeed, de-growth and/or resource redistribution is likely necessary) and that CIR must shape itself to this context.

There is ever-increasing recognition of the issues of sustainability in computing (see for example, the March 2026 special issue of IEEE Software on `Clean Green Software Sustainability') and indeed concrete solutions to improving the environmental sustainability of computing, but efficiency is still frequently prioritised over a consideration of planetary limits and whilst welcome, may not be sufficient. 

CIR does not take place in an organisational vacuum and Silberman emphasises the importance of institutions in addressing environmental problems and that they cannot be ignored or avoided~\cite{silbermanInformationSystemsPractice2023}.  The work we present here embraces this institutional and organisational context as a means to effect change.  Researchers in academia (and increasingly industry) are, in theory at least, subject to ethical oversight.  We focus here on the university/Higher Education Institution (HEI) context and are therefore concerned with bodies such as Research Ethics Committees (RECs) and Institutional Review Boards (IRBs).  National law plays a role in the flexibility and scope that such bodies have (e.g. in the US, IRBs are constituted under the Common Rule with their functions and scope defined in federal regulation~\cite{capiliEthicalResearchInstitutional2024}).  For the remainder of this paper, we will focus on institutional RECs as they are typically constituted in the UK, but the issues may be relevant more widely.  

Knowles et al. identify the need to effect change through high-level leverage points and that educational institutions share part of the joint responsibility for environmentally-cognisant computing advances~\cite{knowlesClimateChangeWhat2025}.  One such potential leverage point is the university research ethics committee.  

Computing researchers are engaged in making value choices when selecting topics of research~\cite{FosteringResponsibleComputing}.  Structurally, RECs are often gatekeepers to the research process and thus have considerable power to increase the scrutiny of choices being made by researchers who use computationally-intensive methods and thereby admit a planetary-limits perspective~\cite{ComputingLimitsCommunications2018} more formally into the ethics oversight and balancing process for all stakeholders.  We acknowledge that there is debate about whether the procedural ethics processes that place RECs in a gate-keeping role are appropriate~\cite{markhamAfterwordEthicsImpact2018} and it is important to mitigate abuse of the gate-keeping power (see Section~\ref{lab:conclusions}).  Nonetheless, whilst it may be desirable to start from a different governance regime (institutional or sector), change can be effected within the status quo (rather than, as can often be the case, relying just on the efforts of concerned individuals as primary actors and agents of change~\cite{revellPromotingSustainabilityProenvironmental2013}), and it is on that ground that we present this paper.  

Institutional RECs typically work within governance frameworks and many universities have policies that govern ethics review.   Researchers may be subject to penalties if they do not obtain favourable opinions or approval in advance of undertaking research.  Crucially, this only applies where the work is within the scope of REC (or equivalent) review.  The scope of an institution's research ethics review policy and the clarity of its inclusion criteria are thus critical in determining whether or not certain types of research receive (or are required to receive) review.  Institutional research ethics policies often acknowledge the environment as an ethically-relevant factor in the scope of review but the level of potential harm required to meet the criteria (and require research to be reviewed) is unspecified.  Thus CIR that does not otherwise trigger ethics review (e.g. by using human data above a level of exemption from review) does not receive scrutiny -- what Ferretti et al. term a `Purview weakness'~\cite{ferrettiEthicsReviewBig2021} and a problem that is also identified by the Data Hazard Labels project~\cite{AVDataHazards}.  Whilst the latter makes the assessment of a broad range of data hazards tractable and straightforward, the institutional review engagement problem remains~\cite{AVDataHazards}.

It is equally important that clear criteria are available to reviewers so that they can make consistent and well-justified comments on the proposed CIR.   Researchers likewise need to be guided on institutional expectations in this respect to facilitate their writing of high-quality applications for ethics review.  There are consequently three main stakeholders to consider: the REC (the body implementing institutional governance in respect of ethics), the ethics reviewer (charged with undertaking balanced and consistent review of harms and benefits), and the researcher (seeking organisational support for their proposed work).

In the context of CIR and in respect of its environmental impacts there are consequently three questions that must be addressed:
\begin{enumerate}
    \item Under what criteria should CIR fall within the scope of research ethics review on the basis only (or at least) of environmental impact considerations?
    \item What criteria should reviewers use to consistently and clearly assess the likely environmental impact of the proposed research to balance this with the expected benefits to society?
    \item How can researchers be supported to reflect on the environmental impacts of their proposed CIR and thereby facilitate the process of review and support?
\end{enumerate}

Answers to these questions must be framed in such a way as to both scrutinise and facilitate research, and the criteria must be realistic and recognise acceptable norms.  In this situation, governance cannot work without the consent and participation of the `governed'; unrealistic criteria may simply be ignored by researchers who do not acknowledge their legitimacy (and will thus not submit their research for review).  As Silberman states: `For the behaviours of individuals in institutional settings to be durably changed, rules must be changed through appropriate processes at the appropriate levels'~\cite[p. 5]{silbermanInformationSystemsPractice2023}.

This paper addresses these questions.  Our aim is not to generate a new ethics framework for CIR, but to draw on those that exist and identify paths to making these practical and operable in the practice of researchers, reviewers and RECs, with the aim of encouraging a greater recognition of planetary limits in CIR and engagement with reflexive practice as a result. 
We first examine the role and practice of RECs in the contemporary UK Higher Education context (Section~\ref{lab:oversight}), and present a small illustrative survey of UK university ethics policies to explore the state of current practice, discussing some of the issues involved (Section~\ref{lab:UKHEI}).  We then take each stakeholder in turn, first addressing the REC and its policy-scope question by proposing some contextually-situated criteria (Section~\ref{lab:scopecriteria}).  Following this, we consider the ethics reviewer, drawing on a range of prior work to propose seven attributes that should be present in any ethics application for CIR (Section~\ref{lab:reviewer}).  In Section~\ref{lab:researcher}, we present a simple approach for researchers to evidence and reflect on their thinking in relation to the attributes through a LIMITS-oriented analysis.  Finally, we identify future work and conclude.  

\section{Research Ethics Governance}
\label{lab:oversight}
Research Ethics Committees (RECs) exist to serve the purpose of independently supervising and balancing the risks and expected benefits of projects proposed by investigators who have an inherent conflict between their duties to protect those involved in their research and the need and desire to obtain scientific results \cite{edwardsRoleRemitFunction2009}.  With the rise of large-scale data use in and for research, the scope of `those involved' has widened considerably over the years and it is not unusual for human-derived data to now be included in ethics review scope as a matter of course alongside human participants themselves.  Environmental concerns are also often now included alongside human involvement but not in ways that are necessarily clear, consistent or actionable (see Section~\ref{lab:UKHEI}).  Among other things, this makes it difficult to appropriately challenge (or even examine) research-benefit justifications for CIR built on narratives of `digital exceptionalism'~\cite{knowlesOurHouseFire2022}.

From a pragmatic standpoint it is important to recognise potential operational difficulties (e.g. what a REC is permitted to consider and comment on - see Section~\ref{lab:conclusions}) and find a way through for the issue at hand.  To avoid some of these difficulties, we locate environmental considerations of CIR securely within the remit of `ethical issues', thereby permitting legitimate commentary by a REC.  It is not our intent to attempt a new framework for environmental issues within research ethics (there is at least one good example of such a framework in the work of Samuel and Richie already~\cite{samuelReimaginingResearchEthics2023} and many examples of criteria of concern across other frameworks and guidance) but instead, our emphasis is on  \emph{operationalising} environmental factors in the ethics review of CIR from a LIMITS perspective.  This means that policies and guidance that aim to capture these issues must be clear and unambiguous in the criteria they apply to ethics review scope and practice.  Since RECs also have a facilitative role in enabling research to proceed, such criteria must not be so broad as to injure this function by imposing additional bureaucracy where it is not required, and also to avoid endangering a REC's social legitimacy among the research community it serves.  Nonetheless, Ferretti et al. identify the importance of re-defining the scope of ethics review to include broader research impact assessments beyond individual interests, noting the implication that less research would be oversight-free~\cite{ferrettiEthicsReviewBig2021}.  Whilst they are writing from a big-data context, the point is valid in relation to environmental impact also.

Thus, from a LIMITS perspective, CIR underpinned by the assumption of continuing exponential growth ~\cite{ComputingLimitsCommunications2018} should fall within ethics review scope, but this must be articulated in such a way as to permit clear identification of that scope (e.g. through screening questions to identify prima-facie risk).  Without such articulation, there is a risk that high-level statements of general research ethics policy about environmental protection become (or remain) disconnected from or denied to REC operations and practice.  

\section{UK Institutional Policy and Practice}
\label{lab:UKHEI}
In this section, we present some observations from a survey of UK HEI research ethics policies and guidance as found on institutional public websites.  This initial non-comprehensive survey was intended to begin exploring the landscape (compared to more extensive surveys in other ethics topics e.g. Poli and Oyebode~\cite{poliResearchEthicsCollaborative2023}); we intend to expand it in future work.   

We selected the Russell Group universities (a self-formed alliance of 24 research-intensive UK institutions~\cite{russellgroup}) on the basis that one would hope and expect to find relevant information in the research ethics policies of such institutions (should it exist anywhere).  The institutional policies were split between the authors, each reading and making notes, and then discussing points of interest noted.  Purves undertook the analysis of UCL's policy (see the Statement of Conflicting Interests).

In each case, we sought research ethics policy documents on the public website of the university concerned.  Where appropriate and available, we also looked at research integrity and misconduct policies as some institutions use these for environmental research issues.  These documents have varying titles and positioning within institutional governance frameworks so are not always directly comparable but we have used our best judgement to classify them appropriately.  In some cases the ethics policies made reference to other institutional guidance and to external frameworks such as research concordats~\cite{integrityconcordat, sustainableconcordat}.  We read the documents for mentions of environmental issues and supported this with a keyword search for `environment', noting any pertinent clauses or other information (additional keywords could be used in an expanded study although in our reading we did not note any widespread use of these).  Since our aim is not to critique specific institutional policies, we have aggregated our observations into themes.

The majority of policies frame ethical risk in relation to human participants and data, though some policies explicitly also note other scenarios in which research should undergo ethics review and it is common for environmental considerations to be included amongst those.  

We found that research ethics policies were available from 22 of the 24 institutions.  Of the remaining two, one policy was behind a staff-only login, and we could not find the other.  The environment is mentioned in 13 of the 22 research ethics policies we found.  Phrasing varies widely and some policies cover different aspects within the same document, thus some policies are counted under more than one theme.  The themes we observed in relation to the environment (in descending frequency) were: impact/implications (8), care and respect (6), risk (5), protection/integrity (4), harm (3), damage (3), and sustainability (2).  One might tentatively observe that less-specific framings occurred more frequently.

Our review of misconduct policies was less comprehensive, taking place primarily where ethics policies linked to them.  Nonetheless, we found that 11 institutions mention the environment in such policies.  Thematically, such mentions can be grouped as: harm to the environment caused by failure to meet obligations (8) failure to protect the environment (3), and consequential (immediate and appropriate action to be taken to prevent further harm) (2).  Of those with misconduct-related environmental statements, we found that in three institutions this was the only place that the environment was mentioned.  We were unable to find ethics or misconduct policy statements relating to the environment for five of the 22 institutions that have research ethics policies.

We noted some guidance and points of interest including a tool for environmental impact investigation at the laboratory level, identification of the need to avoid relying on indirectly harmful practices, and links between research design and respect for natural ecosystems. Finally, and of particular relevance to our theme here, encouragement to use smaller local GenAI models in preference to cloud owing to the environmental impact.

\subsection{Discussion}
In general, relevant comments on environmental considerations seem to be spread over a main ethics or integrity policy, a code of conduct policy and often a misconduct policy. This somewhat scattergun approach may reduce the effectiveness of offering clear guidance to the people who need to operationalise these policies.  The dates around some of the inclusions of these references to the environment and also the way that they cite the two relevant concordats~\cite{integrityconcordat, sustainableconcordat}  suggests these are very recent additions to policies, probably in the last two or three years. Often there is evidence of engagement with the LEAF framework~\cite{LEAFframework} as well - this seems to be a motivator alongside the concordats.  The concordats themselves are broad, high-level institutional statements of commitment.  For example, the Concordat for the Environmental Sustainability of Research and Innovation Practice~\cite{sustainableconcordat} defines six areas of organisational commitment that are in themselves valuable but seem unlikely to engage individual researchers in anything other than general terms.  The Concordat to Support Research Integrity~\cite{integrityconcordat} includes mention of the environment in terms of care and respect, and its expression of research misconduct with respect to the environment is strongly evident in the research misconduct policies of a number of the institutions we surveyed.
 
A minority of institutions have gone further towards providing more concrete advice and guidance or have indicated in action plans that this is forthcoming even though we have not been able to find the latter ourselves. In a couple of cases there are references to the prerogative of the institution to step in and stop a piece of research if there is perceived to be an immediate impact on the environment. There is another policy which notes that there would be concerns about evidence of `short term environmental harm'.  These all point to environmental harm as being something that is defined as immediate or rapid following a piece of research rather than the concept of `slow violence'~\cite{nixonslow} that we are concerned with here (see below).  

The fact that quite a number of these institutions have few references to environmental harm in their main ethics policy but more detail in the misconduct policy is significant because it suggests a `stick before carrot' approach and potentially the pushing back of responsibility and/or consequences onto the individual researchers rather than the institutions or the review committees.  Some of the cited or quoted material in these misconduct policies suggests that they have been obtained from one of the concordats. This focus on misconduct seems significant and interesting, and is somewhat at odds with notions of developing good, reflexive, and positive practices which take these matters into consideration as part of `business as usual' CIR.

Our brief survey indicates that there are a number of barriers to effective and consistent institutional and researcher engagement with the environmental impacts of CIR:
\begin{itemize}
    \item A strong focus on human participation and the use of human data as the defining scope of review policy.
    \item Unclear policies regarding the nature and extent of what is considered to be environmentally harmful, even where this is part of a general policy criterion.
    \item Little or no guidance for researchers on whether their proposed CIR falls within the scope of ethics review in relation to environmental issues, and how to justify it if review is required.
    \item Little or no guidance for ethics reviewers to support balanced and rigorous evaluation of risk and benefit in relation to environmental harm in potential research.
\end{itemize}

Such disconnects between high-level policy and actionable processes are well-recognised, described by Hudson, Hunter and Peckham as the ‘policy-implementation gap’~\citep[p. 1]{hudsonPolicyFailurePolicyimplementation2019}. Kreindler calls the same phenomenon the ‘know-do gap’~\cite[p. 208]{kreindlerWhatIfImplementation2016} and argues that the ‘development of concrete, executable solutions fully informed by understanding of complex, system-level issues [are] the essential bridge between knowledge and action’~\cite[p. 224]{kreindlerWhatIfImplementation2016}. Silberman also notes that there are often differences between what institutions have decided to do and what they are really doing~\cite{silbermanInformationSystemsPractice2023}.
Therefore, whilst we welcome the inclusion of references of environmental harm in research ethics policy, we are concerned that without the development of practical operational guidance, these references risk becoming overlooked, ignored or leading only to superficial acknowledgement, or a sense of cynicism.  
  
The lack of clear scope increases the potential for ethical distancing of researchers from the impact of their research.  An important definition of ethical distancing was given by Kaufmann et al. during their study of cheating by university students, who noted that:

\begin{quote}
`Actors actively seek to increase the distance between themselves and the outcome to rationalize their involvement in unethical behaviour. We refer to this process as ethical distancing because it involves two distinct steps. First, the actor knows that their basic action will result in an unethical outcome. Second, the actor then attempts to create distance between themselves and that unethical outcome.'~\cite[p.128]{kaufmannEthicalDistancingRationalizing2005}
\end{quote}

In the case of RECs, we do not take the view that either applicants or reviewers will commence from the position that a piece of research is inherently `unethical' (although there is something of the prevailing `digital exceptionalism' position ~\cite{knowlesOurHouseFire2022} present in much CIR).  We do suggest that there may nonetheless be aspects of that research inadvertently underpinned by potentially environmentally damaging processes and technologies, for instance within a supply chain or a remote, networked computing service. Due to their arms-length nature from the researchers' local academic context, the practical operation of these processes and technologies may be entirely `opaque' to both researcher and reviewer – an innocent, but all the same `convenient' blind spot from the perspective of ethics review. The concept of ethical distancing helps us consider this possibility, even if it cannot provide an easy solution for overcoming it. Taking a similar view to Kaufmann et al.~\cite{kaufmannEthicalDistancingRationalizing2005}, Shepski notes that `our responsibility for [the wrongdoing of others] varies inversely with our ethical distance from it'~\cite[p. 402]{shepskiGoingEthicalDistance2013} but acknowledges that `there are no firm general rules for deciding which cases are morally acceptable and which are not. In each case, we must bring our intuition, judgment, and the concept of ethical distance to bear' (\emph{ibid.}).

Ethical distance helps identify some of the social, economic and technological dimensions that might potentially cause RECs and researchers to miss some environmental implications of research. In relation to computer science, the concept has been employed in relation to artificial intelligence~\cite{villegas-galavizMoralDistanceAI2024} and automated decision-making~\cite{Democratisingdecisionstechreport} but not, as far as we have been able to identify in relation to environmental concerns. Yet there is also a temporal dimension (congruent with the third-order effects described by Becker et al.~\cite{beckerKarlskronaManifestoSustainability2015}), and here Nixon’s concept of `slow violence' is instructive:
\begin{quote}
`By slow violence I mean a violence that occurs gradually and out of sight, a violence of delayed destruction that is dispersed across time and space, an attritional violence that is typically not viewed as violence at all. Violence is customarily conceived as an event or action that is immediate in time, explosive and spectacular in space, and as erupting into instant sensational visibility. We need, I believe, to engage a different kind of violence, a violence that is neither spectacular nor instantaneous, but rather incremental and accretive, its calamitous repercussions playing out across a range of temporal scales. In so doing, we also need to engage the representational, narrative, and strategic challenges posed by the relative invisibility of slow violence.'~\cite[p. 2]{nixonslow}
\end{quote}

Accordingly, whilst is understandable for a REC to focus on nearer-term ethical consequences of a piece of research consistent with human-level perceptions of cause and effect, this may nonetheless miss more drawn-out forms of environmental (and subsequent social and economic) harm associated with some computationally or materially resource-hungry processes~\cite{leslieEthicsComputationalSocial2023, doggettEnvironmentalClimateJustice2023}.

\section{Stakeholder Perspectives}
In Section~\ref{lab:introduction} we defined three primary stakeholders: the REC (implementing institutional governance), the ethics reviewer (considering whether a researcher has appropriately considered relevant issues), and the researcher (seeking support for their work).  In this section, we examine the problems and perspectives for each.
\subsection{The REC: Policy Scope}
\label{lab:scopecriteria}
For effective and consistent application, we argue that the fit of proposed research to ethics review policy scope should be determined by the answers to yes/no questions.  This mirrors the approach taken to peer review in the ACM SIGSOFT Empirical Standards for Software Engineering~\cite{ralphEmpiricalStandardsSoftware2021}, where criteria are phrased as observable attributes of a paper against which there can be clear and consistent evaluation.  

Ethics policies already have such clauses in relation to the involvement of humans and human data in research (either humans are involved or they are not), but even where environmental impact or harm is mentioned, the criteria are unclear.  What qualifies as environmental harm is typically not specified at the policy level.

We draw on the orders of effects described by Becker et al.~\cite{beckerKarlskronaManifestoSustainability2015} to inform these questions.  Where Becker et al. define first-order effects  as the opportunities and effects created by the existence of software technology and associated processes of production, we translate this to the CIR ethics-scoping context in relation to the local technology resources to be used or developed in the research.  Where second-order effects are defined as the opportunities and effects resulting from the use of software, we translate this to the CIR ethics-scoping context in relation to the technical resources that support the research (e.g., online services).  We defer consideration of third-order effects to the review of an investigation's ethics itself (see Section~\ref{lab:researcher}).

If we are to avoid excessively drawing every investigation using computing technology within scope of ethics review, a threshold must be determined below which research is not considered computationally intensive.  Boundaries are problematic because situations that are close to them are often qualitatively indistinguishable yet are treated differently, and determining a fixed empirical value is both difficult and potentially somewhat arbitrary.  For example, the EU AI Act~\cite{euaiact} determines that GPAI systems pose `systemic risk' if they cumulatively use at least $10^{25}$ floating point operations (FLOPS) in their training.  It is implausible to think that a system trained on fractionally less than this figure is qualitatively less systemic than one trained on slightly more.  In developing the questions below, we have therefore aimed to ground our thresholds in contextually-oriented qualitative criteria rather than absolute values.  By doing so, individual institutions can adapt these where desired.  Whether the contextual thresholds are too conservative or too liberal is perhaps only to be determined in practice.

We propose the following questions for use in screening potential investigations for their inclusion within research ethics review:
\begin{enumerate}
    \item Does the planned research involve specialist hardware (like GPUs or similar specialist units)?  \\\emph{Rationale: the research requires more than `everyday' computing resources, choices about which can affect carbon impact (see ~\cite{grealeyCarbonFootprintBioinformatics2022}) (a first-order effect).}
    
    \item Does the planned research involve cloud computing for reasons of scale or convenience? \\\emph{Rationale: the research is going to use de-materialised (and thus potentially impactfully-opaque) resources to exceed what is possible within the local environment (first and/or second-order effect).}
    
    \item Does the planned research involve algorithms with (broadly) higher computational or data storage requirements than that of typical locally-run desktop software or whose tasks cannot be completed in feasible/reasonable/acceptable time without the use of these? \\
    \emph{Rationale: the research may contribute to the cornucopian  paradigm~\cite{preistUnderstandingMitigatingEffects2016} (first-order effect).}
    
    \item Does the planned research involve services that draw on computing systems that use or rely on any of the above (e.g. LLMs)? \\
    \emph{Rationale: there may be second-order contributions to any environmental harm involved in the research, and it may contribute to third-order effects by building on previous environmentally-impactful research and production.}
\end{enumerate}

If the answer to any question is `yes' then we argue that the proposed research should receive ethics review.  It is important to recall that these initial questions are \emph{not} intended to be answered by, or to elicit, a balanced ethical justification but are intended simply to establish whether a research investigation should go forward into the ethics review process on environmental grounds (in a similar Boolean manner to determining whether it should on the grounds of involving humans or their data).  Questions regarding the ethical balance and justification are presented in the next section.

\subsection{The Ethics Reviewer: Assessing Evidence}
\label{lab:reviewer}
Having established that a proposed investigation falls within the scope of ethics review, the subsequent review must rest on well-justified and consistent areas of consideration in order to be fair to applicants, supportive to reviewers, and secure in terms of institutional responsibilities in law.

\subsubsection{Evidence of Consideration}
\label{lab:evidence}
The ACM SIGSOFT Empirical Standards for Software Engineering ~\cite{ralphEmpiricalStandardsSoftware2021} offer a helpful, attribute-driven model on which to draw (for transparency, we note that the first author of this paper (Gold) contributed to the ethics supplements of these standards).  The standards are focused on expressing the software engineering research community's views on different types of investigation in terms of what should be expected in papers reporting a particular kind of research.  The expectations are characterised in terms of attributes that can be observed or otherwise in papers under review.  These are grouped by: specific attributes, general quality criteria, acceptable deviations, anti-patterns, and invalid criticisms.  Whilst the issues involved in CIR are wider and the requirements of paper peer-review different to what is required in a research ethics application, the model of practical attributes as a set of observable properties is helpful in grounding more abstract considerations of environmental impact, particularly to encourage consistency of approach in ethics review.  We have re-framed the consideration from retrospective (as applied to peer-review of papers) to prospective (as required for the review of a proposed piece of research) but have otherwise adopted a similar expressive approach to that of the SIGSOFT standards~\cite{ralphEmpiricalStandardsSoftware2021}.

We have reviewed a number of ethics guidance frameworks and documents for statements and questions of environmental relevance on which to draw to frame our attribute-style tests for reviewers to apply.  

Samuel and Richie~\cite{samuelReimaginingResearchEthics2023} raise a number of relevant points: data should not be collected or analysed unless outputs will be of sufficient quality; humanity could benefit by prioritising low-technology research especially if the gains will be equal/more than higher technology; data centres should be chosen with sustainability in mind; algorithms should be optimised; data use should be minimised; calculators can be used to determine environmental impact; long-term use of results in terms of carbon expenditure and implications must be considered; mining, manufacturing, and recycling or disposal of physical assets must be considered; and existing technology reused where possible.

The SIGSOFT Empirical Standards Ethics Supplement that relates to Engineering Research~\cite{ralphEmpiricalStandardsSoftware2021} includes a number of criteria for papers reporting this kind of work.  These include describing all plausible and non-trivial potential harms (incorporating environmental destruction and ecological impacts) and attempts to mitigate these; describing the environmental impact of the work in terms of carbon emissions involved in development and use, and other respects relating to materials and mining; and desirable quantification of carbon emissions for development and use of both the artefact concerned and the line of research.

The EU Assessment List for Trustworthy Artificial Intelligence~\cite{EUALTAI} together with earlier associated ethics guidelines~\cite{EUEthicsGuidelines} presents a number of requirements and questions: a requirement for societal and environmental well-being; consideration of the environment and non-human sentient beings as stakeholders in AI systems lifecycles; encouragement toward ecological responsibility and sustainability of AI systems; and ensuring that AI is undertaken with maximal environmental friendliness including supply-chain assessment.  The documents also contain questions for those constructing systems: did you establish ways to measure environmental impact across the system life-cycle, and ensure measures to reduce it?  Are there negative environmental impacts of the system and what are they?  

The UK Government Data and AI Ethics Framework~\cite{DataAIEthics} has a wide range of quite specific considerations for public sector projects that are relevant here.  Briefly summarising these, the areas raised are: whether the proposed approach is the most appropriate technology for the task or whether there are lower-resource or non-compute alternatives that could suffice; whether the minimal amount of data is being used and how this can be further reduced; whether the least storage-expensive data types are to be used; whether the location and timing of data processing has been considered for the lowest ecological impact; and whether analysis has been grounded in metrics of environmental impact where possible (acknowledging the limits of those metrics and mitigating for these if possible).

In the context of bioinformatics, Grealey et al.~\cite{grealeyCarbonFootprintBioinformatics2022} undertake detailed analysis of various algorithms and their associated carbon footprints finding substantial variation that depended on factors including parallelisation, GPU/CPU balance, data-centre energy efficiency, national energy efficiency, and over-allocation of memory.   

Zelenka et al. introduce Data Hazard Labels as a means to enable researchers to characterise and reflectively discuss their research in the context of a controlled vocabulary~\cite{zelenkaDataHazardsOpensource2025}, and show their application to synthetic biology~\cite{zelenkaDataHazardsSynthetic2024}.   A similar discussion is presented by Garc\'{i}a et al.~\cite{garciaDataHazardsEthical2025} in their application of the data hazard labels to computational neuroscience following analysis of the carbon cost of algorithmic choices and assignment of medium-relevance of the High Environmental Impact label.  Baumer and Silberman identify ways to question the appropriateness and necessity of technological solutions~\cite{baumerWhenImplicationNot2011}.  Others (e.g.~\cite{FosteringResponsibleComputing, AIEthicsAIa, floridiAI4PeopleAnEthicalFramework2018, russoGenerativeAISoftware2024}) document broader concerns but have fewer directly adaptable criteria.  

There is a strong emphasis throughout many of these positions for efficiency and lower-power technologies to minimise impact rather than a stronger recognition of planetary limits and the need to consider alternative approaches.  Such concerns are more directly articulated by Blevis~\cite{blevisSustainableInteractionDesign2007}, Preist, Schien and Blevis~\cite{preistUnderstandingMitigatingEffects2016}, Nardi et al.~\cite{ComputingLimitsCommunications2018},  and Knowles et al.~\cite{knowlesClimateChangeWhat2025}. 

\subsubsection{Proposed Evaluation Criteria}
\label{lab:evalcriteria}
The seven criteria proposed here are designed to support an ethics reviewer in consistently assessing whether an application for ethics approval/opinion has considered an appropriate range of environmental issues (and implicitly thus encouraging researchers to do so).  We do not claim that they are novel criteria per se as they draw heavily on the well-understood frameworks and guidelines surveyed above.  However, they are formulated and framed with the specific purpose of being used in ethics review by reviewers rather than applicants.  We conjecture that they cover a sufficiently large number of attributes to effectively assess environmental discussion in CIR applications but few enough to retain engagement and maintain efficient review.  They are intended to enable consistent assessment of the evidence in an application on the basis of clear expectations and justification of the research need.  For example, requiring data minimisation does not imply researchers should use less data than required (or even a small amount of data) but simply that they should consider the data volume actually needed.  It remains for the ethics reviewer to assess the balance of justification against risk in each case (as would always be the situation).  The criteria are not intended to be a reductive checklist; reviewers are free (indeed encouraged) to consider aspects beyond these points and in their interaction.  They may also wish to positively consider how researchers are themselves building ongoing reflexivity into the research methods in areas where it is hard to predict requirements in advance e.g., data minimisation could be assessed more as an attitude and managed risk by the researcher rather than a specific figure to be justified.  This reflects ethical practice in areas such as grounded theory where evolution of ethical issues is expected and a similar balance of protection and flexibility is required~\cite{potrataRethinkingEthicalBoundaries2010a}.  Abbreviations of each criterion (for later use in Section~\ref{lab:researcher}) are shown in parentheses.

\begin{enumerate}
    \item \emph{Data Volume (DV):} The application under review describes how the volume of data to be used will be minimised during initial acquisition, subsequent encoding or processing, and retention.  Opportunities and plans to reduce data storage during the research are clearly identified.

    \item \emph{Necessity (N):} The application under review demonstrates that the technological development or technology use being proposed in the research is necessary and that lower-resource or non-computing approaches could not feasibly achieve the intended outcomes. 

    \item \emph{Storage Duration (SD):} The application under review describes how long data will be retained for, where it will be stored, and for what purposes.  Where data is to be shared (e.g. for open science purposes), a rationale for this is given and weighed against the environmental impact of maintaining the data (even where sharing is a funder requirement).  N.B. this criterion is not intended as an argument against open data sharing, but as a stimulus to ensure that there is a genuine purpose to that sharing and that the location and retention of the data has been considered from an environmental-impact standpoint.

    \item \emph{Compute Requirement (CR):} The application under review demonstrates that the amount of computational effort that will be required does not exceed the minimum required to achieve the intended research outcome.  It justifies the choice of algorithms and their associated environmental costs, describes metrics to measure these during the research, and ways in which deviations from the expected costs will be managed during the research.

    \item \emph{Supply Chain Dependencies (SCD):} The application under review describes any and all supporting services to be used in the research, accounting for their environmental impact and demonstrating attempts made to quantify them e.g., through carbon cost calculations or by appeal to institutional norms, initiatives, or provisions.

    \item \emph{Quantified Impact:} The application under review presents the expected carbon (or other) direct environmental costs of the research using whatever calculators are available and appropriate to the investigation.  It also describes mitigation strategies (e.g. choice of data-centre based on power source or mitigating approaches) and any associated concerns about those strategies (e.g. whether carbon-offsets are appropriate or successful - see Cullenward, Badgley, and Chay~\cite{cullenwardCarbonOffsetsAre2023}).

    \item \emph{Reuse of Technology (RT):} The application under review describes how the research will maximise its use of existing hardware and justifies new acquisitions in terms of their environmental impact and the research necessity (see Blevis~\cite{blevisSustainableInteractionDesign2007}).
    
\end{enumerate}

\subsection{The Researcher: Reflecting on Methods and Impact}
\label{lab:researcher}
Whilst the frameworks surveyed in Section~\ref{lab:reviewer} raise helpful issues for researchers to consider, and our attribute-oriented framing of these for ethics review hopefully brings these into sharp focus for the review process, there is no systematic challenge for researchers to the assumption identified by Nardi et al.~\cite{ComputingLimitsCommunications2018} that exponential computing growth will continue indefinitely.  In this section, we propose a tool to stimulate such reflection in researchers in a systematic way (whilst aiming not to constrain the type or breadth of CIR being undertaken).  Our approach draws on a number of motivations: Markham's `future looking backward' approach to impact ethics~\cite{markhamAfterwordEthicsImpact2018}, Preist, Schien and Blevis' Rubric of Infrastructural Effects (RoIE)~\cite{preistUnderstandingMitigatingEffects2016}, Blevis' Rubric of Material Effects (RoME)~\cite{blevisSustainableInteractionDesign2007}, the visual clarity and present/future considerations of the Data Hazard Labels approach~\cite{zelenkaDataHazardsOpensource2025}, the summary of frameworks expressed in the tests in Section~\ref{lab:evalcriteria}, and planetary limits and the need to challenge  the limitless growth assumption as articulated by Nardi et al.~\cite{ComputingLimitsCommunications2018}.

Markham~\cite{markhamAfterwordEthicsImpact2018} argues for an impact model of ethics to shift thinking away from error-avoidance and concept-driven ethics.  The focus is on future trajectories and longer timescales than the immediate impact of research and is thus congruent with the need identified by Becker et al. ~\cite{beckerKarlskronaManifestoSustainability2015} to consider environmental impact over multiple timescales, the first to third order effects also described there, and congruent also with Nixon's `slow violence' played out over time and space~\cite{nixonslow}.  Markham argues that by taking a speculative future-oriented lens, one can `look back' from the projected future point to the present and consider the trajectory that led to future impact (thereby permitting consideration of unexpected side-effects including in ecology).  

Recalling that our focus here is on CIR under ethics review and that we are focused on environmental impacts, and recognising that this is only part of an overall ethics assessment but one that is currently often under-considered, we need to adapt the broader focus of the frameworks to our specific purpose.

We propose a simple tabular mapping that brings together three key concepts: six of the seven evidential criteria described in Section~\ref{lab:scopecriteria}, concepts extracted from the questions defined by Preist, Schein, and Blevis~\cite{preistUnderstandingMitigatingEffects2016} and Blevis~\cite{blevisSustainableInteractionDesign2007}, and temporal aspects drawing on Markham~\cite{markhamAfterwordEthicsImpact2018}, Preist, Schein, and Blevis~\cite{preistUnderstandingMitigatingEffects2016} and Silberman et al.~\cite{silbermanNextStepsSustainable2014}.  Columns in the table correspond to the evidential criteria (excluding quantified impact which is about how evidence is provided not what it represents) and rows correspond to the impacts extracted from the questions~\cite{preistUnderstandingMitigatingEffects2016,blevisSustainableInteractionDesign2007}.  Within each column are three sub-columns representing the present at year zero (the research itself), and the future at five years and ten years post-research.

We use only six of the seven evidential criteria from~\ref{lab:evalcriteria} since quantification of impact is in itself not an expected cause of growth and is thus excluded.  The remaining six criteria cover the breadth of relevant CIR issues (and are abbreviated to their initials in the table - see the definitions in Section~\ref{lab:reviewer}).  Questions E1 to E5 posed by Preist, Schein, and Blevis~\cite{preistUnderstandingMitigatingEffects2016} are originally framed as balanced questions e.g. `(E1) Does the design encourage infrastructural expansion or obsolescence?'.  We adapt this by extracting the concepts (expansion and obsolescence) and moving the question of encouragement or discouragement into the analysis that populates the table cells.  We approach questions E2 and E3 together (infrastructure use).  We re-cast digital waste (E4) as infrastructural waste, and invert question E5 (sharing of infrastructure) as `infrastructure individualisation' in order that the assessment method of increases being considered harmful is consistent across all criteria.  We also draw out the first principle of Blevis~\cite{blevisSustainableInteractionDesign2007} in relation to material impacts (that invention is linked to disposal), separating the two concepts in the table so that each can be assessed separately on different timescales where required.

As with the attribute statements above, the aim here is to support the elicitation of hidden factors in the balancing decision.  The approach aligns to consideration of the first, second, and third-order effects defined by Becker et al. and the need to play these out over multiple timescales~\cite{beckerKarlskronaManifestoSustainability2015}.  First-order effects of undertaking the research are addressed by year zero, second-order effects are likely to be anticipated at the five-year point, and third-order effects at ten years.  If appropriate, more (or different) time points could be used.
As with the criteria for reviewers, we do not intend this to be a reductive checklist of issues: researchers should engage with the full breadth of issues their work raises.  The framework is intended as a stimulus to reflection, not a limit to it.
\subsubsection{Step One: Cause to Impact Mapping}
Researchers using the map will consider their research design and, from the perspective of each cause, assess the expected impact of the research, its methods, and its outcomes at the years shown and in respect of the environmental-impact drivers in the rows.

Cell values should be either `+', `0', or `-' depending on whether the research and its impact or consequences is likely to have caused (in the case of future years) increase, no change, or to have decreased the impact (row) owing to the cause (column) at the point in time (sub-column).  The use of `have caused' is intentional to reflect Markham's `future-looking-backward' principle~\cite{markhamAfterwordEthicsImpact2018}.  A similar symbolic characterisation is used to capture the relevance of Data Hazard labels to a four-stage model of the research lifecycle~\cite{garciaDataHazardsEthical2025}.

\begin{figure*}[htbp]
    \centering
    \includegraphics[scale=0.6]{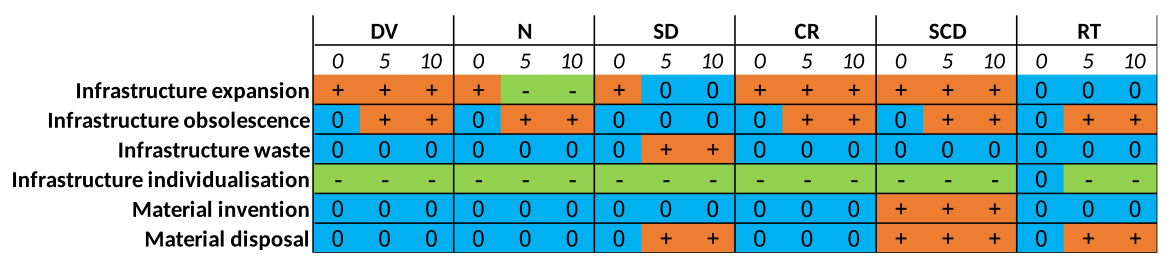}
    \caption{Table showing the first stage of the method: Cause to Impact Mapping.  Columns are the causes, rows are the impact.  Cells show increase (+), decrease (-) or no change (0) at each timescale for each cause-impact pair.}
    \label{fig:example1}
    \Description{A table showing the mapping between causes and impacts at three timescales with the corresponding effect noted as increase, decrease or no change.}
\end{figure*}

An example is shown in Figure~\ref{fig:example1}.  This is a fictitious scenario of research to create a new software engineering tool using third-party LLMs for code generation.  We have `assessed' the project for cause and impact.  Note that the data, compute, and supply chain volumes involved in such activity lead to likely infrastructure expansion both during and after the research.  There are risks in the medium to long-term of infrastructure obsolescence as deployers of any resulting technology are likely to need to upgrade their systems.  We assessed infrastructure waste as relatively benign throughout except in the longer-term for storage duration as it seems likely that the research will produce and/or use a large amount of data that will be retained for replication and/or future use without specific expectations of that reuse.  Infrastructure individualisation seemed likely to be reduced given the use of third-party services shared by others.  

Once the map is completed, the key areas of likely impact and their trajectory from the research forward can be easily identified and particular attention given to these in ethical justification.  Whether taken at a high-level (e.g. the spread or amount of orange highlighting) or in detail, the map offers a simple overview of the research and its expected impact over (in this case) a ten-year timescale.  It also offers clear evidence of consideration for an ethics reviewer.  Naturally, others may assess this differently and in a real scenario more information would be available.  The point is not necessarily that a universally `correct' map is drawn up, but that the concretisation of considerations through the exercise has forced engagement with the issues.

\subsubsection{Step Two: Reflexive Mapping of Controlled Impact to Design}
The tabular map representation also supports an active engagement with the questions posed by Preist, Schien, and Blevis under computing-within-limits and collapse informatics scenarios (questions R1 to R5)~\cite{preistUnderstandingMitigatingEffects2016}.  These are concerned with assumptions of continued growth and whether a service (or in our case, research and its impact) can have value or utility in a reduced or limited resource scenario.  

This can be achieved with the map representation by taking the initial analysis, setting all of the `+' entries in the table to `0', and then evaluating the impact on the six columnar criteria.  In other words, where the initial analysis takes the columns as causes and maps their impact in terms of the impact concepts (rows), this stage inverts the analysis, fixing the impact (rows) to an acceptable level and evaluating the effect of that on the areas of research (columns).

\begin{figure*}[htbp]
    \centering
    \includegraphics[scale=0.6]{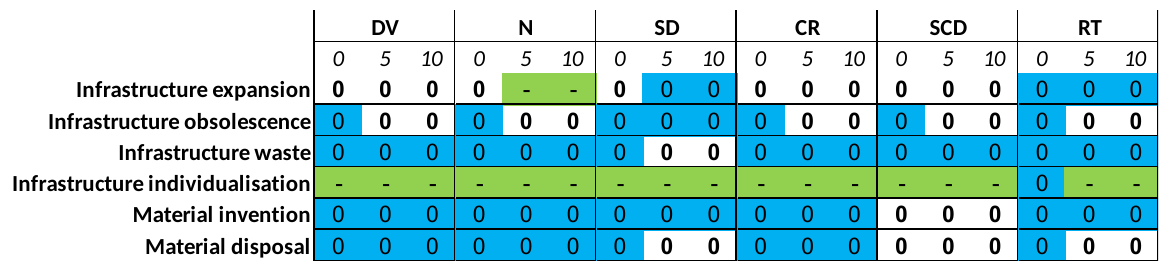}
    \caption{Table showing the second stage of the method: Reflexive Mapping of Controlled Impact to Design.  Impacts (rows) previously identified as growing (+), are fixed to `no change' (0).  Changes to the paired cause in the design of the research can then be reconsidered to achieve this restriction (or if not possible, the growth-dependence is made explicit).}
    \label{fig:example2}
    \Description{A table showing the reflexive second stage of the method in which the values of cells are changed to zero from plus to stimulate thinking about the reverse impact on the research of constraining growth.}
\end{figure*}

This can be seen in Figure~\ref{fig:example2} where the `+' entries from Figure~\ref{fig:example1} have been set to `0'.  This would require a re-evaluation of the feasibility of the research itself in respect of data volume, necessity, compute requirement, and supply-chain dependencies, and of the longer-term value of the research in all six areas.

There are variations on this analysis that could be undertaken depending on the severity of limits being considered.  For example, rather than setting `+' to `0', one could set all `0's to `-' as well.  In an extreme case, one could set every criteria to `-' and assess the research from that perspective.  One could also adopt a prioritised approach, addressing first the most significant impacts (e.g. those pertaining to infrastructure expansion in the example shown), then re-assessing from step 1 under a revised investigation design (to account for knock-on effects of changes made).  This would be a truly reflexive analysis moving back and forth between ethical choices and methodological choices and of the kind encouraged by Markham~\cite{markhamEthicMethodMethod2006}.

Research that simply cannot be undertaken without the `+' entries in the table is thus shown to depend on growth in computing infrastructure and capability in either the short or longer-term and must then grapple explicitly and ethically with the harm that is expected to result.  It will also need to consider the longer-term value of research artefacts produced (data, software etc) if these could not retain their value under a LIMITS scenario.

\subsubsection{Step 3: The Present as the Future of the Past}
Knowles et al~\cite{knowlesClimateChangeWhat2025} noted that there is a lack of impact evaluation in hindsight.  We believe our approach may be able to help with this given its underpinnings in Markham's `future-looking-backward' orientation~\cite{markhamAfterwordEthicsImpact2018}.  This could be achieved by regarding the present (i.e. the research being proposed) as the future of the past.  Thus where our initial characterisation of impact is set at zero, five, and ten years into the future, the analysis could be extended to five and ten years in the past.

It is natural for researchers to present their research as the `next step' in a documented intellectual trajectory.  What we propose here is to also position it as part of an ongoing \emph{environmental} trajectory, not merely the start of a new one.  This will enable an easier assessment of the extent to which the presently proposed research builds on and extends the environmental impact of research that came before it  and thus aid in the assessment of how it will perpetuate, exacerbate, normalise, or even reduce the harmful impacts of past research in the future.  

Researchers may be better able to see how their proposed research (and thus future results) depends on potentially harmful environmental practices and redesign it to mitigate these if needed.  This may help to expose any inadvertent ethical distancing~\cite{shepskiGoingEthicalDistance2013} in play.  By rooting the analysis in past, present, and future, it may be easier for researchers to perceive and acknowledge the impact of their work because it can be characterised in terms of known harms as well as potential ones.

\begin{figure*}[htbp]
    \centering
    \includegraphics[scale=0.6]{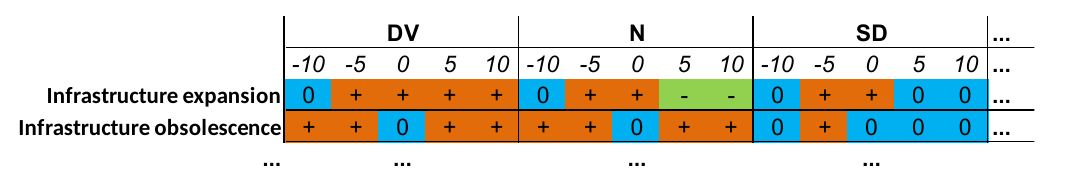}
    \caption{Table showing the third stage of the method: The Present as the Future of the Past.  Additional sub-columns are added to represent the precursors to the research at -5 and -10 years, positioning the current investigation in the centre of a trajectory of enviromental causes and impacts.}
    \label{fig:example3}
    \Description{A table showing the third stage of the method where additional sub-columns have been added to represent five and ten years in the past.}
\end{figure*}

Taking infrastructure expansion in the context of our example above, we can extend the table as shown in Figure~\ref{fig:example3} (for brevity we have only included the first three columns and two rows but this is sufficient to show the principle).  We can see that at -10 years LLMs did not exist (at least not in the form they take now) and thus the environmental trajectory in respect of data volume was benign.  By five years ago the size of the models was evidently increasing and thus the fictitious research we describe here can be seen to be already building on harmful practices in relation to data volume and the proposed investigation would continue this trajectory.  

Some might feel that the consideration of historical emissions or environmental harm as an active ethical process is unnecessarily over-burdensome, perhaps even a form of self-mortification; what is done is, after all, done. It is therefore worth pointing out aspects of academic practice which already consider retrospective perspectives. Most will be familiar with the term ‘standing on the shoulders of giants’ as an expression of debt to previous generations of scholars and a reminder of the importance of critically citing their work~\cite{paulStandingShouldersGiants2018}. Yet since we benefit positively from this legacy we should also be prepared to acknowledge its corresponding, problematic provenance. Such a view, for instance, motivates arguments for researchers to explicitly acknowledge the eugenicist origins of the statistical procedures they employ~\cite{dodsonDisruptingContinuitiesEugenics2024a}. It is now also routine for researchers from many parts of the world to include so-called `land acknowledgements' within presentations, papers and emails which recognise that the land on which their institutions sit was often taken without recompense from indigenous communities~\cite{ lambertRethinkingLandAcknowledgments2021}. Whilst there is always a risk that these kinds of acknowledgements become superficial `virtue signalling', at their best, they present opportunities for `truth-telling, a demand for accountability, and a call to action' (\emph{ibid}, p. 5). We regard our present proposals in this same light.  Moreover, we argue that it presents an opportunity to consider research not just in terms of its intellectual foundations and extension but in its environmental trajectory, assessed at the midpoint between past and future.  In a sense, it allows researchers to unpack in more detail the growth assumption on which their research activity and research outcomes rest.  In turn, this increased requirement to justify the choice of the technology on which research depends will hopefully encourage service providers to become more transparent about their own environmental impact in support of a more sustainable and ethically-justifiable research supply chain.

We have presented the full scope of our approach here but it may be that particular institutions will develop approaches under which the extent of use may be modified in proportion to perceived ethical risk (in much the same way that a full Data Privacy Impact Assessment for the GDPR is not required in all cases).

\section{Conclusions and Future Work}
\label{lab:conclusions}
We have presented three approaches to improving the frequency, consistency, and reflexivity of ethics review of CIR in relation to environmental factors.  Each approach is targeted at a different key stakeholder in the institutional ethics review process: the REC in terms of identifying when a proposal should fall within the scope of review, reviewers in terms of seeking consistent and comprehensive evidence of environmental consideration, and researchers in being supported to undertake reflexive consideration and generate the evidence required.  

Ferretti et al. note that the language of data protection laws has influenced the vocabulary of research ethics~\cite{ferrettiEthicsReviewBig2021}.  In a similar vein, Zelenka et al. aim to support the assessment of data hazards through controlled vocabulary~\cite{zelenkaDataHazardsOpensource2025}.  We hope that having an approach to CIR-related environmental concerns that is designed-for and directly-embeddable within institutional ethics review processes will lead to similarly beneficial outcomes in terms of more frequent and consistent consideration of environmental issues in CIR.  Just as the GDPR has also driven the widespread adoption of data protection training, bringing consistent environmental considerations into ethics review may also beneficially drive carbon literacy training for researchers and reviewers. 

Clear policy and guidance is becoming particularly important for institutional RECs.  In the UK, the Higher Education (Freedom of Speech) Act 2023~\cite{heact} and associated guidance from the Office for Students (OfS) introduced new conditions with which RECs must comply in their operations.  The OfS has responsibility for monitoring this and its guidance, \emph{inter alia}, states~\cite{officeforstudentsRegulatoryAdvice242025}:

\begin{quote}
    `196. Reasonably practicable steps for providers and constituent institutions to take, in relation to research ethics committees, may include:
    \begin{enumerate}
        \item[] a. ensuring that ethical review and requirements are focused on ethical issues and do not impose requirements related to the quality of the proposed research or reputational concerns; \dots'
    \end{enumerate}
\end{quote}

The OfS guidance separates ethical issues from those of research quality but whether this is feasible is open to question (the glossary of terms accompanying the guidance does not define `the quality of the proposed research').  Markham~\cite{markhamEthicMethodMethod2006} argues strongly for a reflexive nature of ethics and methods (linked explicitly to good quality research).  Samuel and Richie likewise link research quality to ethics, indicating that poor quality leads to wasted resources and time~\cite{samuelReimaginingResearchEthics2023}.  Gl{\"a}ser argues the need for a theory of research quality to inform data collection methods in social science research~\cite{glaserHowCanWe2024}.  Since data collection methods carry inherent and varying ethical concerns, quality and ethics are difficult to separate under this argument also.  Margherita et al. undertook a literature-based study with expert validation that found that conformance with ethics was seen as a quality-related attribute of the research process~\cite{margheritaWhatQualityResearch2022}. Velez describes transparency as a `hallmark' of assessing research process quality in data collection, giving the example of checking for ethical considerations when reviewing the methods section of a paper~\cite{velezEnsuringResearchQuality2025}.  In addition, M{\aa}rtensson et al. include ethics within their conceptual model of research quality~\cite{martenssonEvaluatingResearchMultidisciplinary2016} (under the `Conforming' part of their model) thus rendering the two concepts stated in the OfS guidance inseparable under that model.

Without clarity as to what is considered to be in the domain of ethics specifically (as opposed to research quality), the separation of ethical issues and quality of research in the OfS guidance is problematic: the wide space of possible interpretation engenders risk for those involved in ethical scrutiny.

In one sense, the legislation and guidance does not inhibit what should have been (and typically is) the practice of RECs in any case: assessment, commentary, facilitation, and opinion-forming on the ethics of a proposed investigation.  It remains the case that care is needed to be sensitive to the multidisciplinary nature of much contemporary research and to avoid a REC over-shaping research `in its own image' (for example, see Stevenson et al.~\cite{stevensonReconsideringEthicsQuality2015}).  

The legislation and OfS guidance may be motivated by good intent to prevent abuse of the gate-keeping power of RECs and institutions who may be concerned with reputation and/or impose unjustified opposition to particular lines or approaches to research.  However, the risk is that as a result, reviewers and RECs will lack the confidence to critically address aspects of proposed research that they need to.  The 'chilling effect' that the new legislation aims to eliminate from research activity perhaps chills the practice of ethical oversight instead.

Much of this paper has been focused on `pushing' CIR researchers to consider the environment through institutional processes but there is also a role for publishing communities in `pulling' such considerations as expected parts of research papers~\cite{goldetalvalues2022}. Just as Zelenka et al.  suggest the possibility of including data hazard labels within manuscripts~\cite{zelenkaDataHazardsSynthetic2024}, the cause-impact maps we define here could also form part of supplementary material accompanying publications. 

Future work will include expanding the scope and systematicity of the survey of institutional ethics policies to build a more comprehensive picture of the UK HEI sector in terms of environmental considerations in research ethics.  In addition, we plan to undertake a survey of those involved in ethics review and governance within HEIs to research the views, understanding, and considerations of those involved in scrutinising ethics applications.  We also hope to survey CIR researchers for their views.  Finally, we hope to develop collaborations with HEIs to allow us to pilot the framework presented here in practice and in the context of research ethics training.

\section{Ethical Statement}

The empirical research reported herein used publicly-available organisational documents published to university websites.  Consequently it did not require ethics review under UCL's Research Ethics Policy and does not raise issues of ethics relating to human research participants or their data.  It did not qualify as computationally-intensive research under the definition given here.

\section{Statement of Conflicting Interests}
Gold is a UCL ethics reviewer, co-chair of UCL's Computer Science REC and a member of UCL's REC.  He thus has operational responsibilities for implementing UCL's Research Ethics Policy and has some influence on the future direction of ethics policy and  governance for UCL.  He is also a member of the UK Research Integrity Office (UKRIO) expert advisory community.  Purves is an ethics reviewer for the UCL Institute of Education REC.  

The work reported here has not been undertaken at the direction of UCL or UKRIO and not for the specific benefit or use of either organisation. The views expressed here are those of the authors, not their institution.

\begin{acks}
We would like to thank the reviewers for their insightful and helpful comments.
\end{acks}

\bibliographystyle{ACM-Reference-Format}
\bibliography{paper}

\end{document}